\documentclass[reprint,superscriptaddress,amsmath,amssymb,aps,prl]{revtex4-1}

\ifx\pdfoutput\@undefined\usepackage[usenames,dvips]{color}
\else\usepackage[usenames,dvipsnames]{color}
\usepackage{graphicx}% Include figure files
\usepackage{bm}% bold math
\usepackage{amssymb}
\usepackage{hyperref}
\usepackage{color}

\usepackage{amsmath}
\usepackage{eufrak}

\usepackage[T1]{fontenc}
\usepackage[latin1]{inputenc}

\newcommand{\Col}[2]{\left(\begin{array}{c}#1\\#2\end{array}\right)}
\newcommand{\Matrix}[4]{\left(\begin{array}{cc}#1&#2\\#3&#4\end{array}\right)}
\newcommand\dg{^{\dagger}}
\newcommand\w{\omega}
\renewcommand{\vec}[1]{\mathbf{#1}}
\newcommand\rv{\vec{r}}
\newcommand\Ev{\vec{E}}
\newcommand\Hv{\vec{H}}

\newcommand\Evt{\tilde{\vec{E}}}
\newcommand\Hvt{\tilde{\vec{H}}}

\newcommand\Dv{\vec{D}}
\newcommand\Bv{\vec{B}}

\newcommand{\eps}{\varepsilon}
\newcommand{\epst}{\tilde{\eps}}
\newcommand{\mut}{\tilde{\mu}}

\newcommand{\braket}[2]{\left\langle #1 \middle| #2 \right\rangle}
\renewcommand{\Re}{\mathrm{Re}}
\renewcommand{\Im}{\mathrm{Im}}
\newcommand\s{\nu}
\newcommand\nng{\tilde{n}}

\renewcommand{\S}{\mathfrak{S}}

\usepackage{graphicx}% Include figure files
\usepackage{color}

\begin{document}

\title{Electromagnetic helicity in complex media}

\author{F. Alpeggiani}
\affiliation{Department of Quantum Nanoscience, Kavli Institute of Nanoscience Delft, Delft University of Technology, Lorentzweg 1, 2628 CJ Delft, The Netherlands}
\author{K. Y. Bliokh}
\affiliation{Theoretical Quantum Physics Laboratory, RIKEN Cluster for Pioneering Research, Wako-shi, Saitama 351-0198, Japan}
\affiliation{Nonlinear Physics Centre, RSPE, The Australian National University, Canberra, Australia}
\author{F. Nori}
\affiliation{Theoretical Quantum Physics Laboratory, RIKEN Cluster for Pioneering Research, Wako-shi, Saitama 351-0198, Japan}
\affiliation{Physics Department, University of Michigan, Ann Arbor, Michigan 48109-1040, USA}
\author{L. Kuipers}
\affiliation{Department of Quantum Nanoscience, Kavli Institute of Nanoscience Delft, Delft University of Technology, Lorentzweg 1, 2628 CJ Delft, The Netherlands}

%\date{\today}

\begin{abstract}
Optical helicity density is usually discussed for monochromatic electromagnetic fields in {\it free space}. It plays an important role in the interaction with chiral molecules or nanoparticles. Here we introduce the optical helicity density in a {\it dispersive isotropic medium}. Our definition is consistent with biorthogonal Maxwell electromagnetism in optical media, the Brillouin energy density, as well as with the recently-introduced canonical momentum and spin of light in dispersive media. We consider a number of examples, including electromagnetic waves in dielectrics, negative-index materials, and metals, as well as interactions of light in a medium with chiral and magnetoelectric molecules. 
\end{abstract}

\maketitle

%%%%%%%%%%%%%%%%%%%%%%%%%%%%%%%%%%%%%%%%%%
%\section{Introduction}
%%%%%%%%%%%%%%%%%%%%%%%%%%%%%%%%%%%%%%%%%%

{\it Introduction.---}
Helicity is a fundamental property of relativistic spinning particles, such as electrons and photons, which can be regarded as the projection of the spin angular momentum onto the linear momentum direction \cite{messiah}. 
In the case of photons, i.e., electromagnetic fields, helicity is a conserved quantity associated with the dual (electric-magnetic) symmetry of Maxwell's equations \cite{Calkin,Zwanziger,Deser,Gaillard,Stepanovsky,Trueba,Drummond}. 
Recently, studies of electromagnetc helicity got a second wind 
\cite{Tang2010, Hendry2010, Tang2011, Bliokh2011, Choi2012, Schaferling2012, Hendry2012, Barnett1, Corbaton2012, Andrews2012, Torres2012, Barnett2, Corbaton2013, Bliokh2013, Dionne2013, Philbin, Ebbesen2013, Bliokh2014, Nieto,GutscheSPIE, GutscheACS,KruiningGoette},  
because of its close relation to the {\it optical chirality} 
\cite{Barron,Schaferling, Andrews2018}. In particular, it was shown that the circular dichroism in local interactions of light with chiral molecules or nanoparticles is determined by the optical helicity density \cite{Tang2010,Hendry2010,Tang2011,Choi2012,Bliokh2014,Nieto}.

In most previous studies, the helicity density was defined only for monochromatic {\it free-space} optical fields. At the same time, modern nanooptics and photonics often deals with electromagnetic modes in inhomogeneous and dispersive {\it optical media}, including metamaterials and plasmonic nanostructures. Instead of pure photons, there one has to deal with collective light-matter excitations (``quasiparticles''), such as cavity or waveguide photons, plasmons, or optical polaritons. Characterizing the fundamental physical properties of such modes, including their momentum, spin, and helicity, is an essential and rather nontrivial problem (see, e.g., the Abraham-Minkowski momentum controversy \cite{Brevik, Pfeifer, Barnett2010, Milonni, Kemp}). 
In particular, quantifying the helicity of optical fields in complex media is an important task: solving it would allow one to rigorously determine how ``chiral'' the field is. 

The energy density for monochromatic electromagnetic waves in a dispersive optical medium characterized by the permittivity $\varepsilon(\omega)$ and permeability $\mu(\omega)$ was derived by Brillouin almost a century ago \cite{Landau}. 
Recently, some of us obtained analogous expressions for the canonical (Minkowski-type) momentum, spin, and orbital angular momentum in dispersive media \cite{Bliokh2017PRL,Bliokh2017NJP}. However, several recent attempts to introduce the electromagnetic helicity density in optical media \cite{Philbin,Nieto,GutscheSPIE,GutscheACS} faced considerable difficulties. 
Namely, while the energy, momentum, and spin densities for plane waves in homogeneous media are naturally determined by the frequency $\omega$, wavevector ${\bf k}$, and $\sigma {\bf k}/k$ (where $\sigma \in (-1,1)$ is the degree of circular polarization), respectively, the suggested helicity densities do not yield the expected value of $\sigma$ and depend on the medium parameters $\varepsilon$ and $\mu$ \cite{Bliokh2017NJP}. This contradicts the physical picture of the helicity as the spin projection and makes it impossible to compare the chirality of electromagnetic fields in different media.

In this paper, we solve the helicity-in-a-medium puzzle and derive a rigorous expression for the helicity of electromagnetic modes in inhomogeneous, dispersive, and lossless optical media, including dielectrics, negative-index materials, and metals. Our theory is based on the dual-symmetric Hamiltonian formulation of macroscopic electrodynamics \cite{Raghu, Silveirinha2015, Silveirinha2017, Lein} and the identification of the proper helicity operator for the field in macroscopic dispersive media. 
This formalism incorporates in a straightforward way the expressions for the Brillouin energy density, as well as the canonical momentum and spin densities \cite{Bliokh2017PRL,Bliokh2017NJP}. 
Thus, the energy, momentum, spin, and helicity constitute a consistent set of fundamental properties of the electromagnetic modes (akin to those in free-space fields \cite{Bliokh2013,Bliokh2014}), providing a rigorous characterization and deeper understanding of nanophotonic fields.
We illustrate our general results with benchmark examples and discuss the role played by the helicity density in the interaction with chiral and magnetoelectric molecules.

%%%%%%%%%%%%%%%%%%%%%%%%%%%%%%%%%%%%%%%%%%
%\section{Biorthogonal electromagnetism}\label{biorthogonal}
%%%%%%%%%%%%%%%%%%%%%%%%%%%%%%%%%%%%%%%%%%

{\it Biorthogonal electromagnetism.---}
We consider complex amplitudes $\Ev(\rv)$ and $\Hv(\rv)$ of monochromatic electromagnetic fields in a generic dispersive inhomogeneous lossless medium characterized by the real-valued permittivity $\eps(\rv,\w)$ and permeability $\mu(\rv,\w)$. Maxwell's equations can be written as an eigenvalue problem for the electromagnetic ``bispinor'' $\bm{\psi} = (\Ev,\Hv)^T$ \cite{Silveirinha2015, Silveirinha2017, Lein}:
\begin{equation}
\label{maxwell}
\hat{M}^{-1}\! \Matrix{0}{i\nabla\times}{-i\nabla\times}{0} \Col{\Ev}{\Hv} \equiv \hat{\mathcal{H}}\, \bm{\psi} = \w\, \bm{\psi},
\end{equation}
where
\begin{equation}\label{M}
\hat{M}(\rv,\w) = \Matrix{\eps(\rv,\w)}{0}{0}{\mu(\rv,\w)}
\end{equation}
is the constitutive matrix, and throughout the paper we use Gaussian units. 
%with $\eps_0=\mu_0=1$.
We also assume suitable boundary conditions guaranteeing the eigenfrequency $\w$ to be real.
%, such as periodic or ideal conductor boundary conditions, so that each modal eigenfrequency $\w$ is guaranteed to be real.

The ``Hamiltonian'' $\hat{\mathcal{H}}$ in Eq.~\eqref{maxwell} is not Hermitian with respect to the standard bilinear product $\braket{\bm{\psi}_n}{\bm{\psi}_{n'}} = \int d^3\rv \, (\Ev_n^*\cdot\Ev_{n'} + \Hv_n^*\cdot\Hv_{n'})$. 
However, following the prescriptions of biorthogonal quantum mechanics \cite{brody}, one can define a biorthogonal basis of right and left (adjoint) eigenvectors,
which fulfil the biorthogonality condition $\langle\tilde{\bm{\psi}}_n|\bm{\psi}_{n'}\rangle = \delta_{nn'}$. The adjoint eigenvector satisfies $\hat{\mathcal{H}}\dg \tilde{\bm{\psi}} = \w\, \tilde{\bm{\psi}}$, and it is immediate to verify that, for \emph{nondispersive} systems, $\tilde{\bm{\psi}} =  (\eps\Ev,\mu\Hv)^T \equiv (\Dv, \Bv)^T$. In dispersive media, similarly to the Hamiltonian approach introduced in Refs.~\cite{Raghu, Silveirinha2015, Silveirinha2017, Lein}, the adjoint vector reads (see Supplemental Material \cite{SM}):
\begin{equation}
\tilde{\bm{\psi}} \equiv \Col{\Evt}{\Hvt} = \Col{\epst\, \Ev}{\mut\, \Hv}~,
%\Evt(\rv) = \epst(\rv,\w) \Ev(\rv), \quad
%\Hvt(\rv) = \mut(\rv,\w) \Hv(\rv),
\end{equation}
where $\epst(\rv,\w)=\partial [\w\,\eps(\rv,\w)] / \partial \w$ and $\mut(\rv,\w)=\partial [\w\,\mu(\rv,\w)] / \partial \w$.

Since we will deal with local densities of the helicity, energy, momentum, etc., it is instructive to define the local expectation value of an operator $\hat{O}$ as:
\begin{equation}
\label{expectation}
{O} = g\, \Re\! \left( \tilde{\bm{\psi}}^\dag \hat{O} \bm{\psi} \right)
= g\,\Re\! \left(\epst\, \Ev^*\hat{O}\Ev + \mut\,\Hv^*\hat{O}\Hv\right),
\end{equation}
where $g = (16\pi\w)^{-1}$, and we consider the real parts of the local expectation (or weak, with the post-selection in the coordinate eigenstate) values \cite{Bliokh2013weak}. 
%The validity of this formula for the electromagnetic helicity is supported by several physical considerations discussed below. 
Applying this formalism to the operators of energy $\hat{W}=\w$, momentum $\hat{\vec{P}} = -i\nabla$, and spin-1 $\hat{\vec{S}}$ \cite{BB1996, Berry2009, Bliokh2015}, we immediately obtain the Brillouin energy density \cite{Landau}, as well as the canonical momentum and spin densities derived in \cite{Bliokh2017PRL,Bliokh2017NJP}:
\begin{eqnarray}
\label{brillouin}
W & = & g\,\w \left(\epst |\Ev|^2 + \mut |\Hv|^2\right), \nonumber \\
{\bf P} & = & g\, \Im\! \left[\epst\, \Ev^*\!\cdot (\nabla)\Ev + \mut\, \Hv^*\!\cdot (\nabla)\Hv \right], \nonumber \\
{\bf S} & = & g\, \Im\! \left(\epst\, \Ev^*\!\times \Ev + \mut\, \Hv^*\!\times \Hv \right).
\end{eqnarray}
Note that in the biorthogonal formalism, the medium parameters appear explicitly only in the adjoint vector $\tilde{\bm{\psi}}$ (they cannot be symmetrized between the left and right vectors), and therefore these are not subject to the operator action.

%%%%%%%%%%%%%%%%%%%%%%%%%%%%%%%%%%%%%%%%%%
%\section{Electromagnetc helicity}
%%%%%%%%%%%%%%%%%%%%%%%%%%%%%%%%%%%%%%%%%%

{\it Helicity operator and density.---}
Having this general quantum-like formulation of electromagnetism, allowing one to compute the local expectation value of any operator, we need to identify the correct helicity operator. It is useful to express the permittivity and the permeability in terms of the phase refractive index $n$ and dimensionless impedance $Z$ of the medium: 
\begin{equation}
\label{relations}
\eps(\rv,\w) = \frac{n(\rv,\w)}{Z(\rv,\w)}, ~~
\mu(\rv,\w) = n(\rv,\w)\, Z(\rv,\w),
%Z(\rv,\w) = \pm \sqrt{\frac{\mu(\rv,\w)}{\eps(\rv,\w)}},
%\eps(\rv,\w) = \frac{n(\rv,\w)}{c Z(\rv,\w)}, \quad \mu(\rv,\w) = \frac{n(\rv,\w) Z(\rv,\w)}{c},
\end{equation}
i.e., $n=\pm \sqrt{\eps\mu}$, $Z=\pm \sqrt{\mu/\eps}$, where the signs of the square roots are chosen by analytical continuation from the upper half of the complex-frequency plane in agreement with the principle of causality \cite{Smith}. 

We now put forward the helicity operator, which can be written in the following equivalent forms:
\begin{equation}
\label{helicity}
\hat{\S} = \frac{\hat{\vec{S}} \cdot \hat{\vec{P}}}{|n| k_0} 
= \frac{\nabla\times}{|n| k_0} 
= \Matrix{0}{i \s Z}{-i \s Z^{-1}}{0},
\end{equation}
where $k_0 = \w/c$ and $\s(\rv,\w) = n(\rv,\w) / |n(\rv,\w)|$. The first definition in Eq.~(\ref{helicity}) provides the projection of the spin-1 operator onto the momentum direction, assuming the local momentum (wavevector) magnitude $|\hat{\vec{P}}| = |k| = |n| k_0$, and different definitions are equivalent in view of Maxwell's equations (\ref{maxwell}). Note that introducing the absolute value of the refractive index and parameter $\s$ is crucial in the case of {\it negative-index} materials \cite{Veselago,Bliokh2004,Smith2004,Shalaev} ($\eps < 0$, $\mu <0$), where $n<0$, $\nu=-1$, and in {\it metallic} media ($\eps\mu <0$), where $n$ and $Z$ become imaginary, so that $\s = \pm i$. This distinguishes our approach from that in \cite{Corbaton2013}, which is valid only for dielectric dispersionless media. 

The electromagnetic helicity is intimately related to the {\it dual symmetry} between the electric and magnetic fields 
\cite{Calkin, Zwanziger, Deser, Gaillard, Drummond, Barnett1, Corbaton2012, Barnett2, Corbaton2013, Bliokh2013,KruiningGoette}. Namely, the helicity operator provides a generator of the dual transformation (rotation in the ``electric-magnetic'' plane): 
$\bm{\psi}' = \exp(i\theta\hat{\S})\,\bm{\psi}$, where $\theta$ is the real-valued parameter of this transformation. This dual transformation reads:
\begin{equation}
\label{rotation}
\Col{\Ev'}{\Hv'} = \Matrix{\cos(\s\theta)}{- Z \sin(\s\theta)}{Z^{-1} \sin(\s\theta)}{\cos(\s\theta)} \Col{\Ev}{\Hv}.
\end{equation}
Remarkably, in the case of perfect dielectric media ($\s = \pm 1$), Eq.~\eqref{rotation} produces a rotational transformation similar to that in Refs.~\cite{Corbaton2013,KruiningGoette}, while for perfect metals ($\s = \pm i$) it reduces to the {\it hyperbolic} transformation:
\begin{equation}
\label{hyperbolic}
\Col{\Ev'}{\Hv'} = \Matrix{\cosh\theta}{\mp |Z| \sinh\theta}{\mp |Z^{-1}| \sinh\theta}{\cosh\theta} \Col{\Ev}{\Hv}.
\end{equation}
It is the presence of $\s$ in the operator (\ref{helicity}) that makes the corresponding dual transformations (\ref{rotation}) and (\ref{hyperbolic}) \emph{real-valued}. This  guarantees that the transformations preserve their forms in time-dependent Maxwell's equations with real-valued fields.
For a system with a spatially homogeneous impedance, $\nabla (\s Z) = \nabla (\s Z^{-1}) = 0$, it is easy to prove that the transformations (\ref{rotation}) and (\ref{hyperbolic}) leave Maxwell's equations (\ref{maxwell}) invariant. This symmetry of macroscopic Maxwell's equations implies the existence of a conservation law, where the conserved quantity should be identified with the electromagnetic helicity in the medium \cite{Corbaton2013,KruiningGoette} (see Supplemental Material \cite{SM}). 

Substituting the helicity operator (\ref{helicity}) into Eq.~(\ref{expectation}), after some algebra, we derive the optical helicity density in a medium in the following laconic form:
\begin{equation}
\label{density}
\S = 2\,g\, \Re\!\left(\s\, \nng \right) \,\Im\! \left(\Hv^*\!\cdot\Ev \right),
\end{equation}
where $\nng(\rv,\w) = \partial [\w\, n(\rv,\w)] / \partial \w$ is the group refractive index.
Equations (\ref{helicity}) and (\ref{density}) are the central results of this paper.
In vacuum, $\s = \nng = 1$, and the helicity density (\ref{density}) coincides with the known definition for monochromatic free-space fields \cite{Bliokh2013,Bliokh2014,Bliokh2015}. However, in a medium, our definition (\ref{density}) differs considerably from the previous suggestions  \cite{Corbaton2013,Philbin,Nieto,GutscheSPIE,GutscheACS,Bliokh2017NJP} due to the presence of the group-index and the prefactor $\s$. The closest result, which coincides with the helicity density (\ref{density}) in the case of nondispersive dielectric media ($\s =1$, $\nng = n$), was recently obtained in \cite{KruiningGoette}.
Importantly, in the Supplemental Material \cite{SM} we also derive the time-domain expression for the helicity density, the local conservation law (continuity equation) in terms of the time-dependent fields/potentials, and show that the {\it helicity flux} corresponding to Eq.~(\ref{density}) is given, for monochromatic fields, by $\vec{\Sigma} = g \, \Im\! \left(\nu Z^{-1} \Ev^*\! \times \Ev + \nu Z\, \Hv^*\! \times  \Hv \right)$. Thus, in contrast to free-space fields \cite{Barnett1,Barnett2,Bliokh2013}, the helicity flux $\vec{\Sigma}$ differs from the spin density ${\bf S}$ and agrees with the results of \cite{KruiningGoette} for nondispersive dielectric media.
Below we consider the main properties/applications of the helicity density (\ref{density}) and show that our definition provides a consistent and physically meaningful picture of helicity in optical media.

(i) As a measure of chirality, the helicity density (\ref{density}) is even with respect to the time-reversal (${\mathcal T}$) symmetry and odd with respect to the spatial-inversion (${\mathcal P}$) symmetry \cite{Barron}. Therefore, any mirror-symmetric electromagnetic mode  must have zero integral helicity, $\langle \S \rangle =0$.

(ii) In a system with a homogeneous impedance, one can choose the electromagnetic eigenmodes to be also eigenstates of the helicity operator \eqref{helicity}. As we show below, a possible choice of the helicity basis in homogeneous dielectrics is provided by the circularly-polarized plane waves with maximal helicity $\S = \pm W/\w$. When a homogeneous-impedance system is also mirror-symmetric, it is possible to construct both helicity and mirror-symmetry (such as linearly-polarized waves) eigenstates from different linear combinations of the eigenmodes. This explains why such systems possess frequency eigenmodes which are degenerate in pairs.

(iii) In transparent media ($\eps\mu > 0$, $\s = \pm 1$), the helicity density is locally proportional to the group refractive index $\nng$.  Since $\nng > 0$ in passive systems  \cite{Landau}, the sign of the helicity is determined by the sign of the phase index $n$. This implies the inversion of the helicity in negative-index materials, reflecting the inversion of the direction of the canonical momentum (wavevector) with respect to the energy flux (Poynting vector) \cite{Veselago,Bliokh2004,Smith2004,Shalaev}.

(iv) Notably, our formalism allows one to quantify the helicity density even in metallic media ($\eps\mu < 0$).
Assuming $\Im\,n >0$, $\s = i$, the helicity density is locally proportional to $-\Im\, \nng$. For example, in a Drude metal, $\eps(\w) = 1 - \w_p^2/\w^2$, $\mu = 1$, and we obtain
$-\Im\,\nng(\w) = 1/ \Im\,n(\w) = \w/ \sqrt{\w_p^2 - \w^2} > 0$.
%As we show below, in the case of metallic systems the sign of the helicity is also determined by the direction of the local canonical momentum.

%%%%%%%%%%%%%%%%%%%%%%%%%%%%%%%%%%%%%%%%%%
\begin{figure}
\includegraphics[width=0.9\columnwidth]{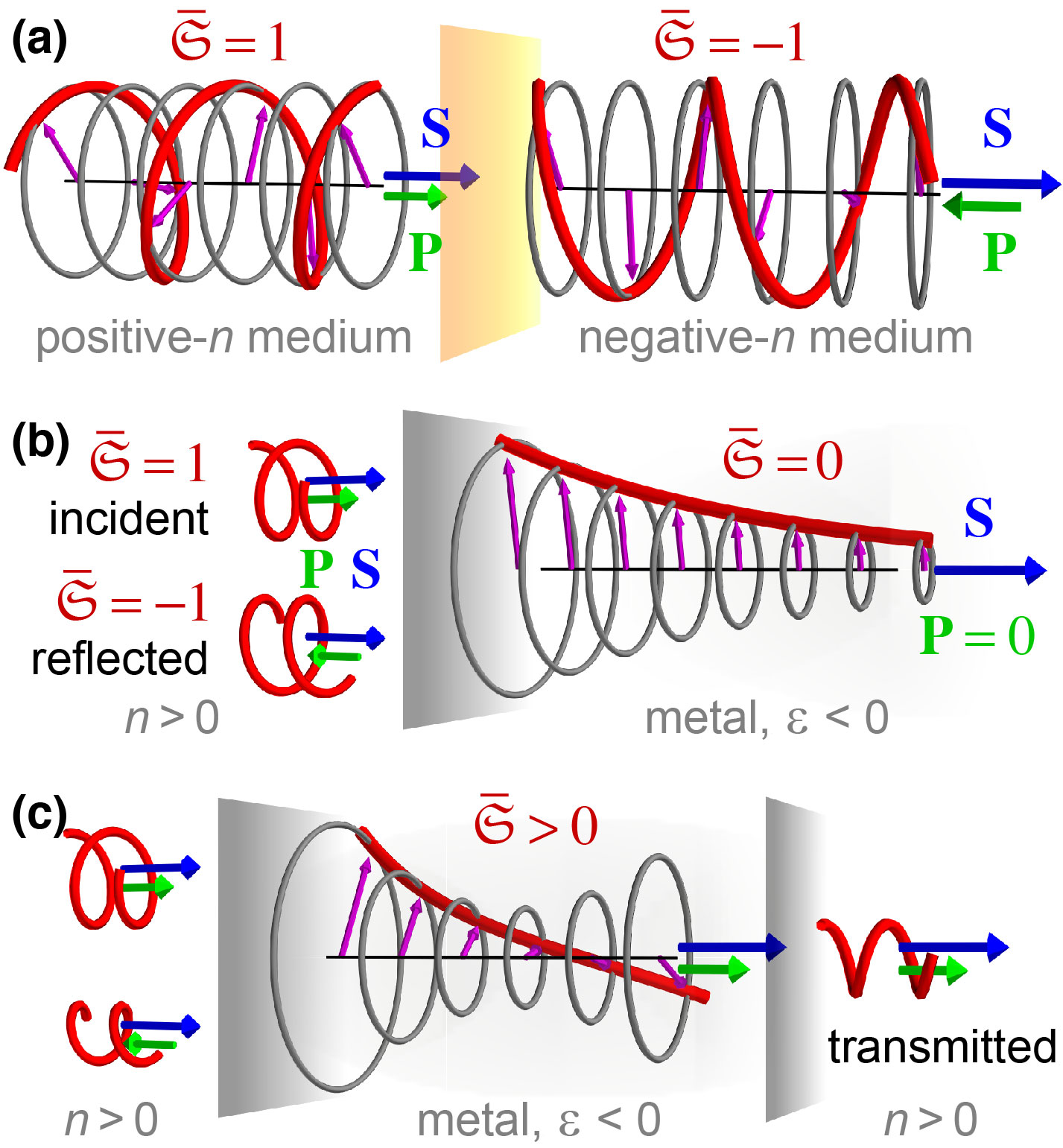}
\caption{Schematic picture of the helicity ($\bar{\S}=\w\,\S/W$), canonical momentum, and spin, when a circularly-polarized plane wave: (a) propagates through an interface between positive-index ($\eps>0$, $\mu>0$) and negative-index ($\eps<0$, $\mu<0$) media; (b) is reflected from a metallic semi-space ($\eps<0$, $\mu>0$); (c) is partially transmitted through a metal film. The red curves and magenta arrows indicate the spatial distribution of the instantaneous electric field $\Re(\Ev\, e^{-i\w t})$, whereas the gray circles show the time evolution of the field at fixed positions. 
\label{scheme}}
\end{figure}
%%%%%%%%%%%%%%%%%%%%%%%%%%%%%%%%%%%%%%%%%%

%%%%%%%%%%%%%%%%%%%%%%%%%%%%%%%%%%%%%%%%%%
%\section{Helicity of plane waves}
%%%%%%%%%%%%%%%%%%%%%%%%%%%%%%%%%%%%%%%%%%

{\it Helicity of plane waves in media.---}
Importantly, our definition \eqref{helicity} provides a meaningful helicity density for plane waves in dispersive media. We first consider a circularly polarized plane wave in a homogeneous transparent medium, $\eps(\w) \mu(\w) >0$, $\s = \pm 1$. Assuming the wavevector 
$\vec{k} =  n k_0 \bar{\vec{z}}$, the electric and magnetic fields read:
\begin{equation}
\label{plane}
\Ev = \frac{A}{\sqrt{2}} 
\left( \begin{array}{c}1\\ i\sigma \\0\end{array} \right) e^{i k_z z}, \quad
\Hv = \frac{-i\sigma}{|Z|}\, \Ev,
%\left[\begin{array}{c}1\\ i\sigma \\0\end{array}\right] e^{i k_z z},
\end{equation}
where $A$ is a constant amplitude, and $\sigma = \pm 1$ determines the circular-polarization sign. Substituting these fields into Eqs.~\eqref{brillouin} and \eqref{density}, and assuming the quantization of energy as $\hbar\w$ per photon, we derive the values of the canonical momentum, spin, and helicity in units of $\hbar$ per photon:
\begin{equation}
\label{ratio_chi}
\frac{\w\, \vec{P}}{ W} = \vec{k}, \quad
\frac{\w\, \vec{S}}{ W} = \sigma\, \bar{\vec{z}}, \quad
\frac{\w\, \S}{ W} = \sigma\, \s.
\end{equation}
These values perfectly correspond to what one can expect for a photon, with helicity $\S = \vec{S}\cdot \vec{P}/ |\vec{P}| = \pm W/\w$. Remarkably, none of the previous approaches \cite{Philbin,Nieto,GutscheSPIE,GutscheACS,Bliokh2017NJP} produced this simple result.
Equations (\ref{ratio_chi}) are written in a form which allows one to consider the transmission of a plane wave from a usual dielectric to a negative-index material, Fig.~\ref{scheme}(a). In such transmission, the momentum and helicity {\it flip their signs}, while the spin does not \cite{Veselago,Bliokh2004,Smith2004,Shalaev,Luo}.

Second, we consider a metallic medium, $\eps(\w) < 0$, $\mu(\w) > 0$, $\s=i$ in the $z>0$ half-space. A circularly-polarized plane wave normally incident on the metal from the vacuum $z<0$ half-space is totally reflected and generates a circularly-polarized {\it purely-evanescent} wave decaying along the $z$-direction inside the metal, Fig.~\ref{scheme}(b). This field inside the metal can be described as a plane wave (\ref{plane}) with a purely-imaginary wavevector $\vec{k} = nk_0 \bar{\vec{z}} = i\kappa\, \bar{\vec{z}}$ and $\Hv = (\sigma / |Z|)\, \Ev$, where $\kappa = \Im(n)\, k_0 > 0$. 
It is easy to show that such {\it non-propagating} wave carries zero canonical momentum and helicity, $\vec{P} = \S =0$, but a nonzero spin given by Eq.~(\ref{ratio_chi}). In agreement with the vanishing helicity, the instantaneous spatial distribution of the electric field is {\it non-chiral}, as shown in Fig.~\ref{scheme}(b).  
Remarkably, considering {\it complex} eigenvalues, not restricted by the real part in Eq.~(\ref{expectation}), brings about the \emph{imaginary} helicity and canonical momentum satisfying the same relations (\ref{ratio_chi}).

To obtain a non-zero real helicity in a metal, one needs to consider a superposition of evanescent waves with opposite decay parameters $\pm \kappa$. Such situation occurs, e.g., in the wave trasmission through a finite-thickness layer of a metal. The corresponding fields are:
\begin{equation}
\label{superposition}
\Ev \!=\! \frac{A e^{-\kappa z}\! +\! B e^{\kappa z}}{\sqrt{2}}
\!\left(\begin{array}{c}1\\ i\sigma \\0\end{array}\right)\!,~
\Hv \!=\! \sigma \frac{A e^{-\kappa z} \!-\! B e^{\kappa z}}{\sqrt{2}\,|Z|}\!
\left(\begin{array}{c}1\\ i\sigma \\0\end{array}\right)\!.
\end{equation}
For these superposition fields, the ratio of the helicity and energy densities \eqref{brillouin} and \eqref{density} is non-zero and equals:
\begin{equation}
\label{helicity_metal}
\frac{\w\, \S}{W}=\sigma\, 
%\frac{\epst |Z|^2 - \mut}{\epst |Z|^2+\mut}\,
\frac{2\, \alpha\, {\rm Im}(A^* B)}{(|A|^2 e^{-2\kappa z} + |B|^2 e^{2\kappa z})
+2\,\alpha \, \Re(A^* B)},
\end{equation}
where $\alpha = (\epst |Z|^2 - \mut)/(\epst |Z|^2+\mut)= - \Im(\tilde{n})|Z|/(\w |n| \partial |Z|/\partial\w)$. Calculating the canonical momentum and spin densities for the fields (\ref{superposition}), we find the following compact relations:
\begin{equation}
\label{helicity_metal_2}
\frac{\w\, \vec{S}}{W}= \sigma\, \bar{\vec{z}}, \quad
{\S} = \sigma\, \frac{\vec{P}\cdot \bar{\vec{z}}}{\kappa} .
\end{equation}
The last equation here reveals the close relation between the helicity and the {\it propagation} of the wave. Figure~\ref{scheme}(c) shows the instantaneous electric-field distribution for the superposition (\ref{superposition}). In contrast to the non-chiral non-propagating field in Fig.~\ref{scheme}(c), this distribution is {\it chiral}, which agrees with its non-zero helicity (\ref{helicity_metal}) and (\ref{helicity_metal_2}).

Thus, in the above examples, the helicity density in a circularly-polarized plane wave corresponds to $\hbar$ per photon (in absolute value), whereas its behaviour reflects fundamental connections with the canonical momentum, spin, and chirality of the field.

%%%%%%%%%%%%%%%%%%%%%%%%%%%%%%%%%%%%%%%%%%
%\section{Interaction with chiral and magnetoelectric matter}
%%%%%%%%%%%%%%%%%%%%%%%%%%%%%%%%%%%%%%%%%%

{\it Interaction with chiral and magnetoelectric matter.---}
One of the main applications of the optical helicity is the probing of chiral or magnetoelectric matter \cite{Tang2010, Hendry2010, Tang2011, Schaferling2012, Hendry2012, Dionne2013, Ebbesen2013, Bliokh2014,Nieto,Fisher,Andrews2018}. So far, only interactions of free-space light with chiral molecules or nanoparticles have been considered. Here we consider the interaction with an admixture of chiral/magnetoelectric molecules in an isotropic optical medium. This is described by the modified constitutive matrix in Eq.~\eqref{M}:
\begin{equation}
\hat{M'} = \hat{M} + \Matrix{\Delta\eps}{i\Delta\xi + \Delta\zeta}{-i\Delta\xi + \Delta\zeta}{0},
\end{equation}
where $\Delta\eps(\rv)$ represents the perturbation of the permittivity (the permeability is not perturbed in practically relevant situations), whereas $\Delta\xi(\rv)$ and $\Delta\zeta(\rv)$ account for the chiral and the magnetoelectric response of the medium, respectively. The magnetoelectric response (also called ``false chirality'' \cite{Barron,Barron2}) is the ${\mathcal P}$-odd and ${\mathcal T}$-odd phenomenon predicted by Curie and Debye \cite{Curie,Debye} and currently attracting considerable attention in electromagnetism and condensed-matter physics \cite{ME1,ME2,ME3,Sihvola,Tretyakov,Hehl,Bliokh2014,Fisher}.

We now consider a propagating electromagnetic mode (either in a homogeneous medium or in a waveguide), which is characterized by the group velocity $\tilde{v}(\w) = \partial\w / \partial k$ and the corresponding modal group index $\tilde{n}^{\text{(m)}}(\w) = c/\tilde{v}(\w)$ ($\tilde{n}^{\text{(m)}} = \tilde{n}$ in a homogeneous medium). 
By treating $\Delta\xi$ and $\Delta\zeta$ as perturbations and using the biorthogonal formalism described above, we calculate the corresponding {\it phase shifts} experienced by light travelling over a distance $L$:
\begin{eqnarray}
\label{dphi}
\frac{\Delta \phi_{\xi}}{k_0 L} = - \frac{2g}{N}
\int_V d^3\rv\, \Re(\Delta\xi)\, \tilde{n}^{\text{(m)}} \Im(\Hv^*\!\cdot\Ev),\\\label{dphi2}
\frac{\Delta \phi_{\zeta}}{k_0 L} = - \frac{2g}{N} 
\int_V d^3\rv\, \Re(\Delta\zeta)\,\tilde{n}^{\text{(m)}}\Re(\Hv^*\!\cdot\Ev).
\end{eqnarray}
Here, $V$ is the volume under consideration and $N = \w^{-1} \int_V d^3\rv\, W(\rv)$ is the normalization factor for the mode. Notably, the same formulas with the substitution $\Re(\Delta\xi)\to \Im(\Delta\xi)$ and $\Re(\Delta\zeta)\to \Im(\Delta\zeta)$ provide the variations of the {\it attenuation coefficient} for the mode, $\Delta {\mathcal{A}}_{\xi,\zeta}$.

By comparison with Eq.~\eqref{density}, it is evident that the chiral phase shift is determined by the local helicity density $\S(\rv)$ with the ``slow-down'' factor $\eta(\rv) = \tilde{n}^{\text{(m)}} / \Re[\nu\tilde{n}(\rv)]$, which accounts for the difference between the group velocity of the mode and the local group velocity in the medium. At the same time, the magnetoelectric response is determined by the {\it imaginary} part of the complex expectation value of the helicity, which can also be called the {\it magnetoelectric density} \cite{Bliokh2014}.

When the chiral/magnetoelectric molecules are localized around a point $\rv_0$, the phase shifts (\ref{dphi}) become proportional to the local helicity and magnetoelectric densities: e.g., $\Delta\phi_{\xi}(\rv_0) = - (k_0 L / N) \eta(\rv_0) \S(\rv_0)$. Note also that the {\it relative} chiral and magnetoelectric responses, introduced in Refs.~\cite{Tang2010,Bliokh2014} for free-space fields, are obtained as a ratio of the phase shifts (\ref{dphi}) with respect to the shift induced by the perturbation of the permittivity, $\Delta\eps$:
\begin{equation}
\frac{\Delta\phi_{\xi}}{\Delta\phi_{\eps}} = \frac{\Delta\xi(\rv_0)}{\Delta\eps(\rv_0)}\,
\frac{2\Im (\Hv^*\!\cdot\Ev)}{|\Ev|^2}.
\end{equation}
Thus, our approach generalizes the results of Refs.~\cite{Tang2010,Bliokh2014} for the case of complex optical media.
It is important to note, though, that the absolute phase shift $\Delta\phi_{\xi}$ depends only on the helicity density.
In this manner, the helicity density $\S$ essentially quantifies the interaction with chiral matter. Our approach allows optimizing the helicity by means of optical media, including engineered nanostructures, which is a viable route for enhancing the optical sensitivity to chiral/magnetoelectric molecules. For example, our results show that circularly-polarized evanescent waves in metals cannot sense chiral inclusions, as their real helicity vanishes. This shows that optical chirality in complex media is not rigidly connected to circular polarization. At the same time, such evanescent waves in metals possess purely imaginary helicity, $\Re(\Hv^*\!\cdot \Ev)=(\sigma/|Z|)\, |\Ev|^2$, thereby offering a perfect tool for probing the magnetoelectric response.

%%%%%%%%%%%%%%%%%%%%%%%%%%%%%%%%%%%%%%%%%%
%\section{Conclusions}
%%%%%%%%%%%%%%%%%%%%%%%%%%%%%%%%%%%%%%%%%%

{\it Conclusions.--}
We have derived the electromagnetic helicity operator and density, which is physically meaningful in dispersive inhomogeneous (but isotropic and lossless) media, including negative-index materials and metals.
This quantity completes the set of dynamical properties of light in optical media, including the Brillouin energy density, canonical momentum and spin \cite{Bliokh2017PRL,Bliokh2017NJP}. We have considered  nontrivial examples of the optical helicity in transparent media and perfect metals, as well as its manifestation in the optical interactions with chiral and magnetoelectric molecules immersed in the medium. Our results can also be applied to systems with small losses/gain, by considering only the real part of the permittivities and permeabilities, similarly to the case of the Brillouin energy density \cite{Jackson}. However, the extension of the present formalism to highly lossy systems is an open task, because of the ambiguity in defining cycle averages with non-oscillatory fields.

%%%%%%%%%%%%%%%%%%%%%%%%%%%%%%%%%%%%%%%%%%
%\section*{Acknowledgements}\label{sec:Acknowledgements}

\vspace*{0.2cm}

\begin{acknowledgments}
This work was supported by 
Marie Sk\l{}odowska-Curie individual fellowship BISTRO-LIGHT (No. 748950), 
the European Research Council (ERC Advanced Grant No. 340438-CONSTANS),
MURI Center for Dynamic Magneto-Optics via the AFOSR Award No. FA9550-14-1-0040, 
the IMPACT program of JST, CREST Grant No. JPMJCR1676, 
JSPS-RFBR Grant No. 17-52-50023, 
the Sir John Templeton Foundation, 
the RIKEN-AIST Challenge Research Fund, 
and the Australian Research Council.
\end{acknowledgments}
%%%%%%%%%%%%%%%%%%%%%%%%%%%%%%%%%%%%%%%%%%

%\bibliographystyle{apsrev4-1}
\bibliography{bib}

\end{document}